\documentclass[aps,prx,superscriptaddress,reprint,longbibliography]{revtex4-1}

\usepackage{graphicx} 
\usepackage{textcomp} 
\usepackage{hyperref} 
\usepackage{ifthen} 
\usepackage{xifthen} 
\usepackage{color} 
\usepackage{amsmath} 
\usepackage{lineno}
\usepackage[english]{babel}
\usepackage{ucs}
\usepackage[utf8x]{inputenc}
\usepackage{amssymb}
\usepackage{amsfonts}
\usepackage{bm}
\usepackage{bbm}
\usepackage{soul}  
\definecolor{hlyellow}{rgb}{0.97,0.88,0.63}
\sethlcolor{hlyellow}

\newcommand{\mm}[1]{\mathrm{#1}}
\newcommand{\ui}{\mathrm{i}}
\newcommand{\ub}{\mathrm{b}}
\newcommand{\up}{\mathrm{p}}
\newcommand{\abs}[1]{\left|#1\right|}
\newcommand{\fig}[2][]{%
	\ifthenelse{\isempty{#1}}
	{Fig.~\ref{#2}}
	{Fig.~\ref{#2}#1}
}

\begin{document}

\title{Classical St\"uckelberg interferometry of a nanomechanical two-mode system}

\author{Maximilian\,J. Seitner}
\email{maximilian.seitner@uni-konstanz.de}
\affiliation{Departement of Physics, University of Konstanz, 78457 Konstanz, Germany}
\affiliation{Center for NanoScience (CeNS) and Fakult\"at f\"ur Physik, Ludwig-Maximilians-Universit\"at, Geschwister-Scholl-Platz 1,
M\"unchen 80539, Germany}
\author{Hugo Ribeiro}
\affiliation{Department of Physics, McGill University, Montreal, Quebec, H3A 2T8, Canada}
\author{Johannes K\"olbl}
\affiliation{Departement of Physics, University of Konstanz, 78457 Konstanz, Germany}
\author{Thomas Faust}
\author{J\"org P. Kotthaus}
\affiliation{Center for NanoScience (CeNS) and Fakult\"at f\"ur Physik, Ludwig-Maximilians-Universit\"at, Geschwister-Scholl-Platz 1,
M\"unchen 80539, Germany}
\author{Eva M. Weig}
\affiliation{Departement of Physics, University of Konstanz, 78457 Konstanz, Germany}
\affiliation{Center for NanoScience (CeNS) and Fakult\"at f\"ur Physik, Ludwig-Maximilians-Universit\"at, Geschwister-Scholl-Platz 1,
M\"unchen 80539, Germany}
\begin{abstract}
The transition from classical to quantum mechanics has intrigued scientists in the past and
remains one of the most fundamental conceptual challenges in state-of-the-art physics.
Beyond the quantum mechanical correspondence principle, quantum-classical analogies have
attracted considerable interest. In this work, we present classical two-mode interference for a nanomechanical two-mode system, realizing classical St\"uckelberg interferometry. In the past, St\"uckelberg interferometry has been investigated exclusively in quantum mechanical two-level systems. Here, we experimentally demonstrate a classical analog of St\"uckelberg interferometry taking advantage of coherent energy exchange between two-strongly coupled, high quality factor nanomechanical resonator modes. Furthermore, we provide an exact theoretical solution for the double passage St\"uckelberg problem which reveals the analogy of the return probabilities in the quantum mechanical and the classical version of the problem. This result qualifies classical two-mode systems at large as a testbed for quantum mechanical interferometry.
\end{abstract}
\maketitle

\section{Introduction}
\label{intro}
In 1932, St\"uckelberg\,\cite{1932_Stuckelberg} investigated the dynamics of a quantum two-level system undergoing a double passage through an avoided crossing. For a given energy splitting, an interference pattern arises that depends
on the transit time and the rate at which the energy
of the system is changed. This discovery lead to the advent of St\"uckelberg interferometry that
allows for characterizing the parameters of a two-level system or for achieving quantum
control over the system\,\cite{2010_Shevchenko_Review}. St\"uckelberg interferometry has been intensively studied in a variety of quantum systems, e.g., Rydberg atoms\,\cite{1992_Yoakum_PRL}, ultracold atoms and molecules\,\cite{2007_Mark_PRL},
dopants\,\cite{2013_Dupont_Ferrier_PRL}, nanomagnets\,\cite{2000_Wernsdorfer_EPL}, quantum
dots\,\cite{2010_Petta_Science,Gaudreau2012,2013_Ribeiro_PRL, 2014_Forster_PRL} and superconducting
qubits\,\cite{2005_Oliver_Science,2006_Silanpaa_PRL,2009_LaHaye_Nature,2012_Shevchenko_PRB,2016_Gong_APL} as well as theoretically in a semi-classical optomechanical approach\,\cite{2010_Heinrich_PRA}. Here, we
experimentally study a classical analog of St\"uckelberg interferometry, the coherent energy exchange of two strongly coupled classical high Q nanomechanical resonator modes, which can be seen as two high occupancy phonon states. We employ the analytical
solution\,\cite{1996_Vitanov_PRA} of the Landau-Zener problem describing the single passage through the
avoided crossing\,\cite{1932_Landau,1932_Zener,1932_Stuckelberg,1932_Majorana} to analyze the
St\"uckelberg problem, demonstrating that the classical coherent exchange of energy follows
the same dynamics as the coherent tunnelling of a quantum mechanical two-level system.\\
The past years have seen the advent of highly versatile nanomechanical systems based on strongly coupled, high quality factor
modes\,\cite{2013_Okamoto_NatPhys,2013_Faust_NatPhys,2014_Shkarin_PRL}. The
strong coupling generates a pronounced avoided crossing of the classical mechanical modes realizing a nanomechanical two-mode system that can be employed as a testbed for the dynamics at energy level crossings\,\cite{2013_Okamoto_NatPhys,2013_Faust_NatPhys,2014_Shkarin_PRL,2012_Faust_PRL}.\\ In the case of a quantum two-level system, e.g. a spin-1/2, a single passage through the avoided crossing results in Landau-Zener dynamics originating from the tunnelling of a quantum mechanical excitation between two quantum states\,\cite{1932_Landau}.
In the classical case, the exchange of excitation energy between two strongly coupled mechanical
modes represents a well established analogy to this process\,\cite{1988_Maris_AJP,2009_Shore_AJP}.
From a quantum mechanical point of view, the two classical modes can be described as high occupancy
states with billions of phonons residing in the respective
resonator mode, where the discrete bosonic energy levels are thermally smeared out orders of
magnitude larger than their level spacing.\\
During a double passage through the avoided crossing
within the coherence time of the system, phase is accumulated, leading to
self-interference. This interference results in oscillations of the return probability, in a quantum mechanical context well-known as
St\"uckelberg oscillations\,\cite{1932_Stuckelberg}, which have previously been studied in many
quantum systems\,\cite{2010_Petta_Science,Gaudreau2012,2013_Ribeiro_PRL,2016_Gong_APL,2011_Sun_PRB}. In the classical case, the
return probability is analogous to the probability that the excitation, namely oscillation energy,
coherently returns to the same mechanical mode. We experimentally demonstrate that phenomenon and present an exact
theoretical solution of the classical St\"uckelberg problem, which demonstrates that the classical flow in
the coherent classical system follows the same dynamics as the unitary evolution operator of a
quantum mechanical two-level system. In this way, classical St\"uckelberg interferometry
opens up a path to further investigate the transition from quantum to classical systems as
has recently been demonstrated in the framework of ultracold
atom experiments\,\cite{2016_Lohse_NatPhys,2016_Kaufman_arxiv,2016_Neuzner_NatPhot}.\\
\section{Nanomechanical two-mode system}
\label{experiment}
\begin{figure}[!htb]
\includegraphics{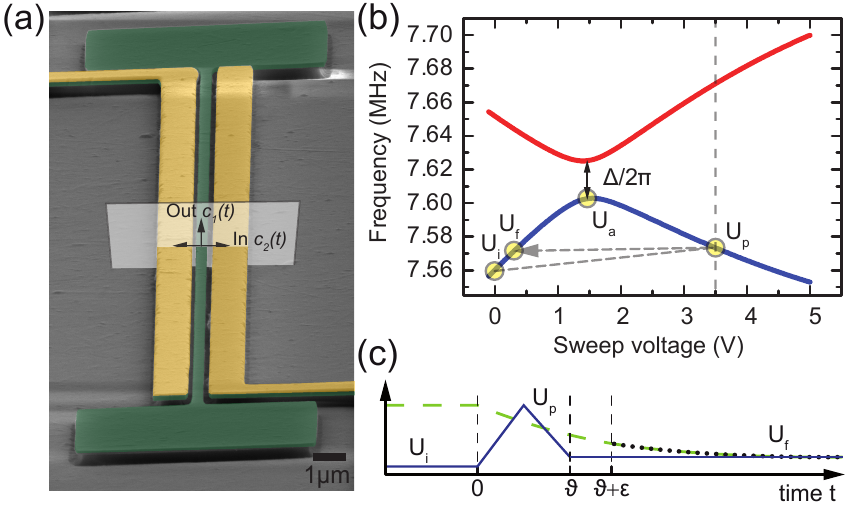}
\caption{\label{fig1}Nanomechanical resonator and measurement scheme. (a) False color scanning electron micrograph of the 50\,\textmu m long, 270\,nm wide and 100\,nm thick silicon nitride string (green) flanked by two adjacent gold electrodes in oblique view. Arrows indicate the flexural mode polarizations out-of-plane (Out) and in-plane (In). The normalized amplitude of the respective mode is denoted $c_1(t)$ for out-of-plane polarization and $c_2(t)$ for in-plane polarization as explained in the text. (b) Avoided mode crossing of sample\,A exhibiting a frequency splitting of $\Delta/2\pi=22.614$\,kHz at the avoided crossing voltage $U_\mathrm{a}=U_\mathrm{i}+1.471$\,V$=9.371$\,V. The lower (out-of-plane) mode is excited at frequency $\omega_\mathrm{1}(U_\mathrm{i})/2\pi=7.560$\,MHz defined by the initialization voltage $U_\mathrm{i}$. An additional sweep voltage applies a triangular voltage ramp rising to a maximum of $U_\mathrm{p}$ and back to the read-out voltage $U_\mathrm{f}=U_\mathrm{i}+0.2$\,V$=8.1$\,V, thus transgressing the avoided crossing twice. The sweep voltage also decouples the mode from the fixed-frequency drive, consequently inducing an exponential decay of the amplitude. (c) Time evolution of the sweep voltage beginning at $t=0$, increasing to $U_\mathrm{p}$ and returning to $U_\mathrm{f}$ after interval $\vartheta$.  The mechanical signal power (green dashed line) is measured after a delay $\varepsilon$ and a fit (black dotted line) is used to extract its magnitude at time $t=\vartheta$. The measured return signal is normalized to the mechanical signal power at $t=0$.}
\end{figure}

We experimentally explore a purely classical, mechanical two-mode system, consisting of two orthogonally polarized
fundamental flexural modes of a nanomechanical resonator (Fig.\,\ref{fig1}\,(a)). The flexural modes belong to the in-plane
and out-of-plane vibration of a 50\,\textmu m long, 270\,nm wide and 100\,nm thick doubly clamped,
high-stress silicon nitride (SiN) string resonator. Dielectric drive and control via electric gradient
fields\,\cite{2012_Rieger_APL} as well as the microwave cavity enhanced, heterodyne dielectric
detection scheme\,\cite{2012_Faust_NatComm,2012_Rieger_APL,2013_Faust_NatPhys} is provided via two
adjacent gold electrodes as detailed in appendix\,\ref{sec:nems}. Applying a dc voltage to the electrodes induces an electric polarization
in the silicon nitride string, which, in turn, couples to the electric field gradient resulting in a
quadratic resonance frequency shift with the applied voltage\,\cite{2012_Rieger_APL}. The electric
field gradients along the in- and out-of-plane direction have opposing signs, and hence an
inverse tuning behavior. Whereas the out-of-plane oscillation shifts to higher mechanical resonant
frequencies, the in-plane oscillation decreases in frequency with the applied dc
voltage\,\cite{2012_Rieger_APL}.  Hence, the inherent frequency offset of in-plane and out-of-plane
oscillation, induced by the rectangular cross-section of the string, can be compensated.
Furthermore, the applied inhomogeneous electric field induces a strong coupling between the two
modes\,\cite{2012_Faust_PRL}. Near resonance, they hybridize into normal
modes\,\cite{2013_Faust_NatPhys}, diagonally polarized along $\pm 45$\textdegree. A pronounced avoided
crossing with level splitting $\Delta/2\pi$ reflects the strong mutual coupling of the flexural
mechanical modes as depicted in Fig.\,\ref{fig1}\,(b).  In order to study St\"uckelberg interferometry, we
perform a double passage through the avoided crossing using a fast triangular voltage ramp. We
initialize the system at voltage $U_i$ in the lower branch of the avoided crossing via a resonant
sinusoidal drive tone at the resonance frequency $\omega_\mathrm{1}(U_\mathrm{i})/2\pi$ of the out-of-plane
oscillation (cf. Fig.\,\ref{fig1}\,(b)). As illustrated in Fig.\,\ref{fig1}\,(c), at time $t=0$, a fast triangular voltage ramp with
voltage sweep rate $\beta$ up to the peak voltage $U_\mathrm{p}$, and back to the read-out voltage
$U_\mathrm{f}$ is applied to tune the system through the avoided crossing. Note that the ramp detunes the system 
from the resonant drive and the mechanical energy starts to decay exponentially. At
$U_\mathrm{f}$, we measure the exponential decay of the mechanical oscillation in the
lower branch after time $t=\vartheta+\varepsilon$, where $\vartheta$ is the duration of the ramp, i.e.
the propagation time, and $\varepsilon$ serves as temporal offset to avoid transient effects. The
signal is extrapolated and evaluated at time $\vartheta$ by an exponential fit and normalized to
the signal intensity at the initialization point ($t=0$), consequently yielding a normalized squared
return amplitude. The return signal has to be measured at the read-out voltage $U_\mathrm{f}$ at $\omega_1(U_\mathrm{f})/2\pi$ since the fixed rf drive tone at $\omega_1(U_\mathrm{i})/2\pi$
cannot be turned off during the measurement. The presented voltage sequence is analogous to the one
employed in Ref.\,\cite{2011_Sun_PRB} and differs from the frequently performed periodic driving
scheme in St\"uckelberg interferometry experiments\,\cite{2010_Shevchenko_Review}.
\section{Finite time St\"uckelberg theory}
\label{theory}
We follow the work of Novotny\,\cite{2010_Novotny_AJP} to derive the classical flow (Hamiltonian flow\,\cite{Arnold_Math})
describing the dynamics of the system in the vicinity of the avoided crossing. We start with
Newton's equation of motion for the displacement
\begin{equation}
	\begin{aligned}
		m \ddot{u}_1 (t) &= -k_1 u_1 (t) -\kappa \left[u_1 (t) - u_2 (t)\right],\\
		m \ddot{u}_2 (t) &= -k_2 u_2 (t) + \kappa \left[u_1 (t) - u_2 (t)\right],
	\end{aligned}
	\label{eq:newton}
\end{equation}
with $u_j (t)$ ($j=1,2$) describing, respectively, the out-of-plane ($j=1$) and in-plane ($j=2$)
displacement of the center of mass of the oscillator, $k_j$ is the spring constant of mode $j$,
$\kappa$ the coupling constant between the two modes and $m$ is the effective mass of the oscillator. We look
for solutions of the form $u_j (t) = c_j (t) \exp[i \omega_1 t]$ with $c_j (t)$ a normalized
amplitude, i.e. $\abs{c_1 (t)}^2 + \abs{c_2 (t)}^2 =1$. In the experimentally relevant limit where 
$\kappa/k_1 \ll 1$, the amplitudes $c_j (t)$ are slowly varying in time as compared to the
oscillatory function $\exp[i \omega_1 t]$. As a consequence, it is possible to neglect the second
derivates $\ddot{c}_j (t)$ in the equations describing the motion of $c_j (t)$, which are
obtained by replacing the ansatz for $u_j (t)$ in Eq.~\eqref{eq:newton}. Thus, the system of coupled
differential equations describing the evolution of the normalized amplitudes is
\begin{equation}
	\begin{cases}
		i\dot{c}_1 = \frac{\kappa}{2 \omega_1 m } c_2,\\
		i\dot{c}_2 = \frac{\kappa}{2 \omega_1 m}c_1 -
		\frac{(\omega_2-\omega_1)^2}{2 \omega_1}  c_2 ,
	\end{cases}
	\label{eq:amp_c1c2}
\end{equation}
with $\omega_\mathrm{j} = \sqrt{k_j/m}$ the bare resonance frequency of mode $j$ in units of $2\pi$.
In the vicinity of the avoided crossing, where the modes can exchange energy, we have $\omega_2\simeq \omega_1$ such that
$(\omega_2-\omega_1)^2/ 2\omega_1 \simeq \omega_2-\omega_1$. If we
further assume $\omega_2 - \omega_1 \simeq \alpha t$, with $\alpha$ the frequency sweep rate, and
define $\Delta=\left|\lambda\right|= \kappa/(m \omega_1)$, Eq.~\eqref{eq:amp_c1c2} reduces to
\begin{equation}
	i \dot{\mathbf{c}}(t) = H (t) \mathbf{c} (t),
	\label{eq:c1c2_matrix}
\end{equation}
with $\mathbf{c} (t) = (c_1 (t)\,c_2(t))^{\mm{T}}$ and
\begin{equation}
	H(t) =
	\begin{pmatrix}
		0 && \frac{\lambda}{2}\\
		\frac{\lambda}{2} && -\alpha t 
	\end{pmatrix}.
	\label{eq:dynmat}
\end{equation}
Since we are interested in multiple passages through the avoided crossing, we look for the classical
flow $\varphi(t,t_\ui)$ defining the state of the system at time $t$ given that we know its state at
some prior time $t_\ui$, $\mathbf{c} (t) = \varphi(t,t_\ui) \mathbf{c} (t_\ui)$. Typically, $\mathbf{c}
(t_\ui)$ is the initial condition of the system. One can show that the classical flow obeys the
same differential equation as $\mathbf{c} (t)$, $i \dot{\varphi}(t,t_\ui) = H(t) \varphi(t,t_\ui)$.
By applying the time-dependent unitary transformation $S(t) =\exp[i\alpha t^2/4]\mathbbm{1}_2$ to
the classical flow, i.e. $\varphi(t,t_\ui) = S(t) \tilde{\varphi}(t,t_\ui) S^\dag(t_\ui)$, we find that
$\tilde{\varphi}(t,t_\ui)$ obeys the differential equation,
\begin{equation}
\begin{aligned}
	i \dot{\tilde{\varphi}} (t,t_\ui) &= \left(S^\dag (t) H(t) S(t) -i S^\dag (t) \dot{S} (t)\right)
	\tilde{\varphi} (t,t_\ui)\\
	 &= \tilde{H}(t) \tilde{\varphi} (t,t_\ui),
	\label{eq:classicalflowLZSM}
	\end{aligned}
\end{equation}
with 
\begin{equation}
	\tilde{H} (t) = 
	\begin{pmatrix}
		\frac{\alpha t}{2} && \frac{\lambda}{2}\\
		\frac{\lambda}{2} && -\frac{\alpha t}{2}
	\end{pmatrix},
	\label{eq:c1c2_matrix_lzsm}
\end{equation}
where $\mathbbm{1}_2$ denotes the unity operator in two dimensions. Equation~\eqref{eq:classicalflowLZSM} coincides with the Schr\"odinger equation for the unitary evolution operator of the Landau-Zener problem with $\hbar$ set to 1, for which an exact
finite-time solution is known\,\cite{1996_Vitanov_PRA} (see also appendix\,\ref{sec:theo}). With the help of the classical flow, one can easily calculate the state of the system
after a double passage through the avoided crossing (St\"uckelberg problem). We find
\begin{equation}
	\mathbf{c} (t) = \varphi_\ub (t, -t_\up) \varphi (t_\up, t_\ui) \mathbf{c}(t_\ui),
	\label{eq:solstueckelberg}
\end{equation}
with $\varphi_\ub (t, t_\ui) = \sigma_x \varphi (t, t_\ui) \sigma_x$ describing the evolution of the
system during the back sweep (see appendix\,\ref{sec:theo}) where $\sigma_x$ denotes the Pauli matrix in x-direction and $t_\up$ labels the time at which
the forward (backward) sweep stops (starts). From Eq.~\eqref{eq:solstueckelberg}, one can obtain the return probability
to mode 1,
\begin{equation}
\begin{aligned}
	&P_{1\to 1} =\\
	 &\abs{\varphi_{11} (t_\up,t_\ui) \varphi_{11}^\ast(t, -t_\up) +
	\varphi_{12}^\ast (t_\up,t_\ui) \varphi_{12}^\ast(t, -t_\up)}^2,
	\label{eq:returnprob}
	\end{aligned}
\end{equation}
with $\varphi_{ij} (t,t_\ui)$ the matrix elements of $\varphi (t,t_\ui)$. Note that we use the frequency sweep
rate $\alpha$ in the theory, which is converted to the experimentally accessible voltage sweep rate
$\beta$ via a conversion factor $\zeta=55.042$\,kHz/V as elucidated in appendix\,\ref{sec:conv}.\\
The analogy between the unitary evolution operator and the classical flow, both expressed in the
basis of uncoupled states (modes), allows one to draw the analogy to the quantum mechanical return
probability in St\"uckelberg interferometry. In principle, this corresponds to the averaging over all
possible Fock states in the phonon distribution of the mechanical resonator mode. The normalized amplitudes are associated with the normalized energy in each resonator mode and differ conceptually from the probability that a quantum mechanical two-level
system is found in either of the two quantum states. Nevertheless, the dynamics of the normalized
amplitudes in classical St\"uckelberg interferometry is analogous to the dynamics of the quantum
mechanical probabilities in the sense that the coherent exchange of oscillation energy between two
coupled modes can be associated with the transfer of population between two quantum states. A more detailed
discussion and comparison of our theoretical approach to previous models\,
\cite{1988_Maris_AJP,1996_Vitanov_PRA,2009_Shore_AJP,2010_Shevchenko_Review} will be published
elsewhere\,\cite{2016_Seitner_arxiv}.\\
Note that St\"uckelberg interferometry does in the contrary not apply to the case of two coupled quantum harmonic oscillators in a general quantum state. In that case, the effective model describing the dynamics would
resemble that of the multiple-crossings Landau-Zener problem\,\cite{1997_Usuki_PRB}, which leads to a much
more complex dynamics than the standard St\"uckelberg dynamics. Only in the case of a singly populated quantum level, i.e., the single phonon Fock state, and if the Hamiltonian of the system conserves the number of excitations, the discussed result is recovered.
\section{Classical St\"uckelberg interferometry}
\label{exp}
\begin{figure*}[!htb]
\includegraphics{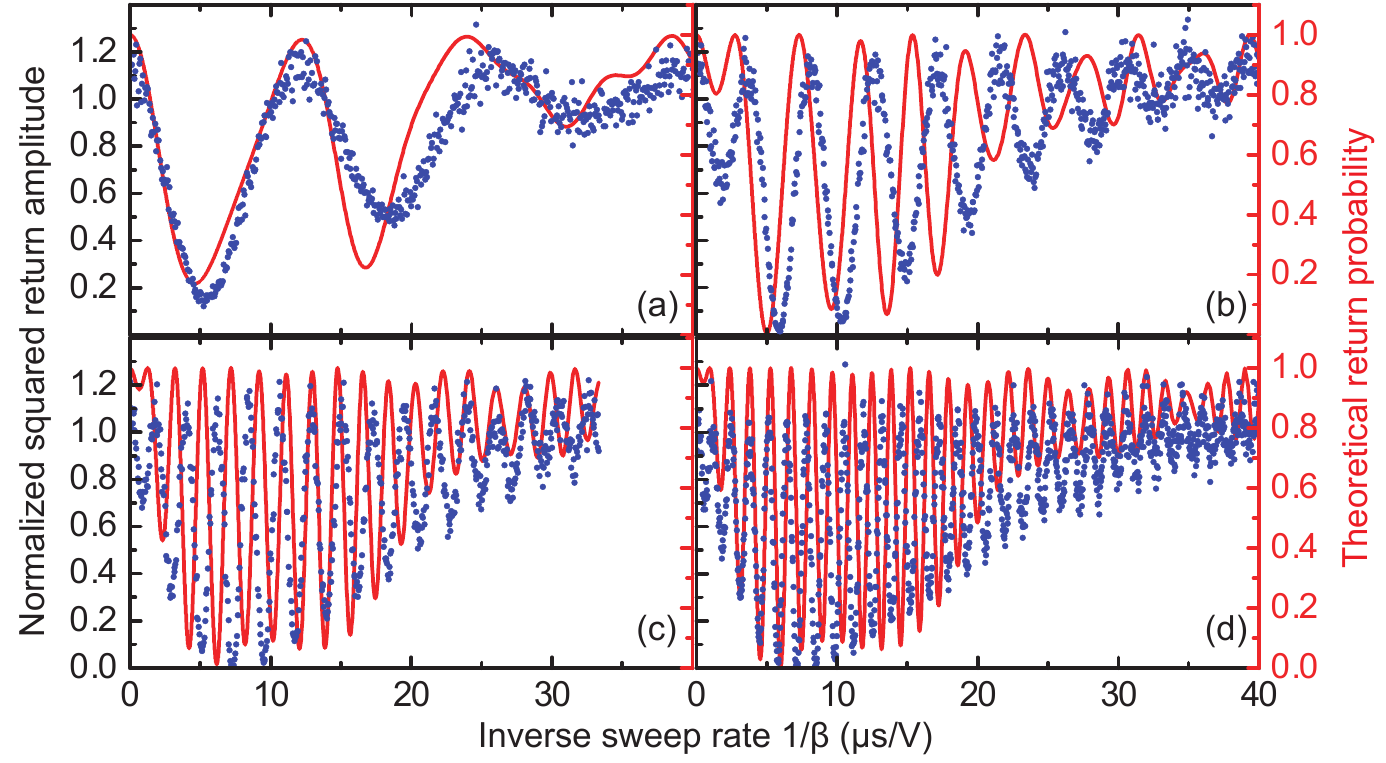}
\caption{\label{fig2}Classical St\"uckelberg oscillations. Normalized squared return amplitude (left axis, blue dots) and theoretically calculated return probability (right axis, red line) versus inverse sweep rate for fixed peak voltages of $U_\mathrm{p}=2.5$\,V\,(a), $U_\mathrm{p}=3.5$\,V\,(b), $U_\mathrm{p}=4.5$\,V\,(c) and $U_\mathrm{p}=5.0$\,V\,(d) measured on sample\,A.}
\end{figure*}
Experimentally, we investigate classical St\"uckelberg oscillations with two different samples in a 
vacuum of $\leq$ $10^{-4}$\,mbar. Sample\,A is investigated at 10\,K in a
temperature-stabilized pulse tube cryostat which offers a greatly enhanced stability of the
electromechanical system against temperature fluctuations. Sample\,B is explored
at room temperature in order to confirm the results and to check their stability under ambient temperature fluctuations. Note that in both experiments, the system operates deeply in the classical
regime\,\cite{2013_Faust_NatPhys} and does not exhibit any quantum mechanical properties. Sample\,A exhibits a mechanical quality factor
$Q=\omega/\Gamma \approx 2 \times 10^5$ and linewidth $\Gamma/2\pi \approx 40$\,Hz at
resonance frequency $\omega_{1}(U_\mathrm{i})/2\pi=7.560$\,MHz of the 50\,\textmu m long string
resonator ensuring classical coherence times in the millisecond regime\,\cite{2013_Faust_NatPhys}. The
level splitting $\Delta /2\pi =22.614$\,kHz exceeds the mechanical linewidth by almost three orders
of magnitude, which puts the system deep into the strong coupling regime. We initialize the system
at $U_\mathrm{i}=7.9$\,V and apply triangular voltage ramps with different voltage sweep rates
$\beta$ for a set of peak voltages $U_\mathrm{p}$. Figure\,\ref{fig2} depicts the normalized squared return
amplitude for different peak voltages and the theoretical return probabilities calculated without
any free parameters. The normalized squared return amplitude may exceed a value of unity due to
normalization artefacts which arise from the different signal magnitudes at the initialization and
read-out voltages in addition to measurement errors.  We observe clear oscillations in the return
signal in good agreement with the theoretical predictions for lower peak voltages. As the number of
oscillations increases for higher peak voltages, the deviation from the theoretical prediction is
more pronounced. We attribute this to uncertainties and fluctuations of the characteristic sweep
parameters of the system, which change under application of the voltage ramp and over time as
discussed in appendices\,\ref{sec:temp} and \ref{sec:fluc}. A further deviation arises from the assumption that a
linear change of the voltage leads to a linear change of the difference in frequency. This
is only an approximation since the mechanical resonance frequencies tune quadratically
with the applied voltage\,\cite{2012_Rieger_APL}. However, since most of the energy exchange
happens in the vicinity of the avoided crossing, where the difference in frequency is
linearized, one expects to see noticeable deviations from theory only for higher peak
voltages.\\
\begin{figure*}[!htb]
\includegraphics{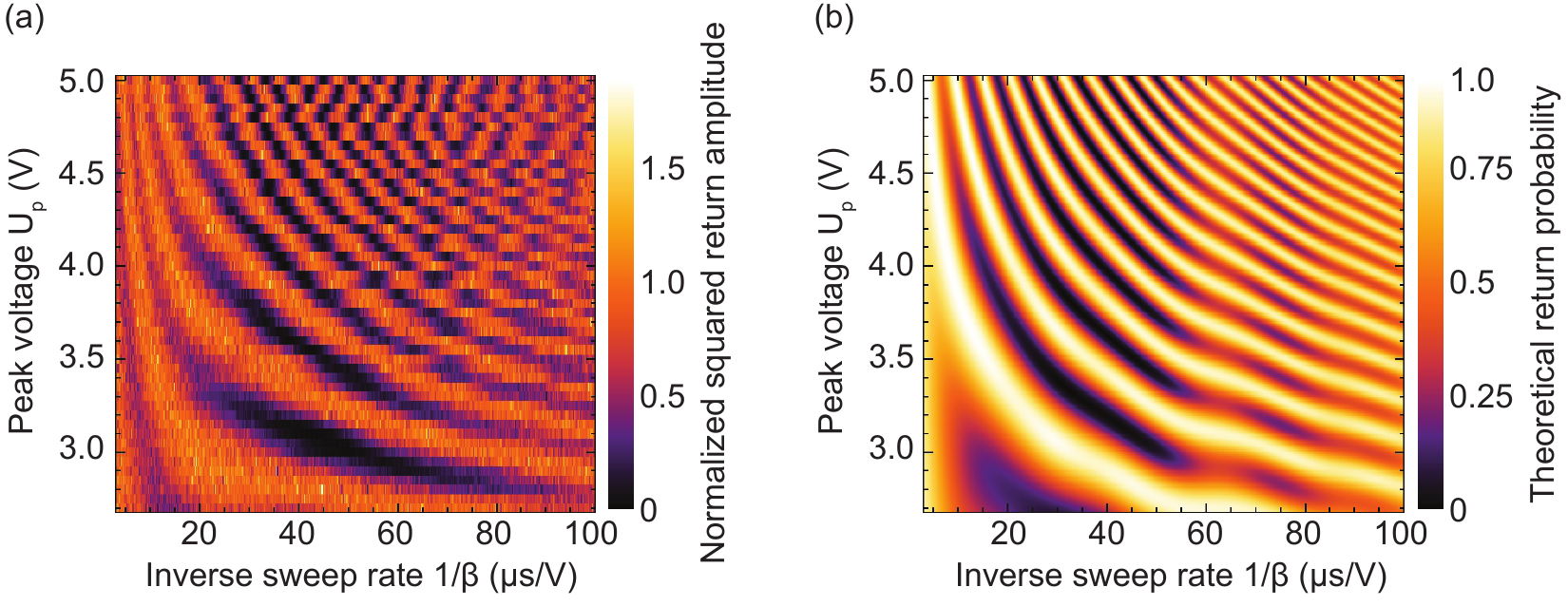}
\caption{\label{fig3}Comparison of experimental data and the theoretical model. (a) Color-coded normalized squared return amplitude versus inverse sweep rate and peak voltage measured on sample\,B. The dataset is not interpolated. (b) Color-coded theoretical return probability given by Eq.~\eqref{eq:cj} versus inverse sweep rate and peak voltage for the equivalent data range. The theory is calculated with a single set of parameters, extracted from the avoided crossing illustrated in appendix\,\ref{sec:conv} (Fig.\,\ref{conv}) and contains no free parameters.}
\end{figure*}
In order to reproduce the experimental data and to test the stability of classical St\"uckelberg
interferometry against fluctuations, we repeat the experiment on a second sample of the
same design at room temperature (sample\,B, denoted by index "B"). The now 55\,\textmu m long resonator has
a mechanical linewidth of $\Gamma_\mathrm{B}/2\pi \approx 25$\,Hz at frequency
$\omega_\mathrm{B,1}(U_\mathrm{B,i})/2\pi=6.561$\,MHz, which results in a quality factor of
$Q_\mathrm{B}\approx 2.6\times 10^5$ at the initialization voltage $U_\mathrm{B,i}=10.4$\,V and
hence an improved mechanical lifetime of 6.21\,ms. Furthermore, the sample exhibits a mode
splitting of $\Delta_\mathrm{B}/2\pi=6.322$\,kHz and a conversion factor of
$\zeta_\mathrm{B}=19.224$\,kHz/V. Figure\,\ref{fig3} depicts a color-coded two-dimensional map of the
normalized squared return amplitude as a function of the inverse voltage sweep rate $\beta$ and the
peak voltage $U_\mathrm{p}$ alongside the theoretical return probability of the classical
St\"uckelberg oscillations, again calculated with no free parameters. We investigate double passages up to a total propagation time of $\vartheta=1.0$\,ms
in the experiments conducted on sample\,B. To account for the decay
of both modes when tuned away from the drive for the considerably longer ramps applied to sample\,B, we model the
mechanical damping by an exponential decay with an averaged decay time $t_0 = 5.7\,\mm{ms}$. After a
measurement time $t_m$, the probability to measure an excitation of mode $j$ is given by
\begin{equation}
\label{eq:cj}
\abs{c_j(t_m)}^2 = \exp[-t_m /t_0] P_{1\to j},
\end{equation}
with $P_{1\to 1}$ given by Eq.~\eqref{eq:returnprob} and
$P_{1\to 2} = 1- P_{1\to 1}$. The experimental data shows remarkably good agreement with the
theoretical predictions, despite temperature fluctuations of several kelvin per hour, which shift
the mechanical resonance frequency up to 40 linewidths. In order to initialize the system at the
same resonance frequency in each measurement, a feedback loop regulates the initialization voltage
$U_\mathrm{i}$ (see appendix\,\ref{sec:temp}). Consequently, the recording of a single
horizontal scan at a fixed peak voltage in Fig.\,\ref{fig3}\,(a) takes up to 16 hours, incorporating a
non-negligible amount of fluctuations of the system parameters, such as e.g. the center voltage of
the avoided crossing $U_\mathrm{a}$, which imposes considerable uncertainties on the parameters used
for the theoretical calculations. To further illustrate the influence of fluctuations, Fig.\,\ref{fig4}\,(a) and
Fig.\,\ref{fig4}\,(b) depict horizontal and vertical line-cuts of the two-dimensional map in Fig.\,\ref{fig3} at
$U_\mathrm{p}=3.3$\,V and at inverse sweep rate $1/\beta=51.6$\,\textmu s/V, respectively.  For
small inverse sweep rates, i.e. very fast sweeps, the experimental data in Fig.\,\ref{fig4}\,(a) deviates from
the theoretical model due to a flattening of the voltage ramps in the room temperature experiment
(see appendix\,\ref{sec:fluc}). For sweeps with $1/\beta\geq 50$\,\textmu s/V,
the experimentally observed St\"uckelberg oscillations exhibit good agreement with the theoretical
predictions even for the line-cut along the vertical peak voltage axis (cf. Fig.\,\ref{fig4}\,(b)). Note that
Fig.\,\ref{fig4} depicts the best results from all datasets at room temperature. Further exemplary line-cuts
are provided in appendix\,\ref{sec:fluc}, also exhibiting a clear oscillatory behavior
in the normalized squared return amplitude, but incorporating larger deviations from theory in
certain regions and therefore revealing fluctuations of system parameters over time, predominately
induced by temperature drifts.\\
\begin{figure}[!htb]
\includegraphics{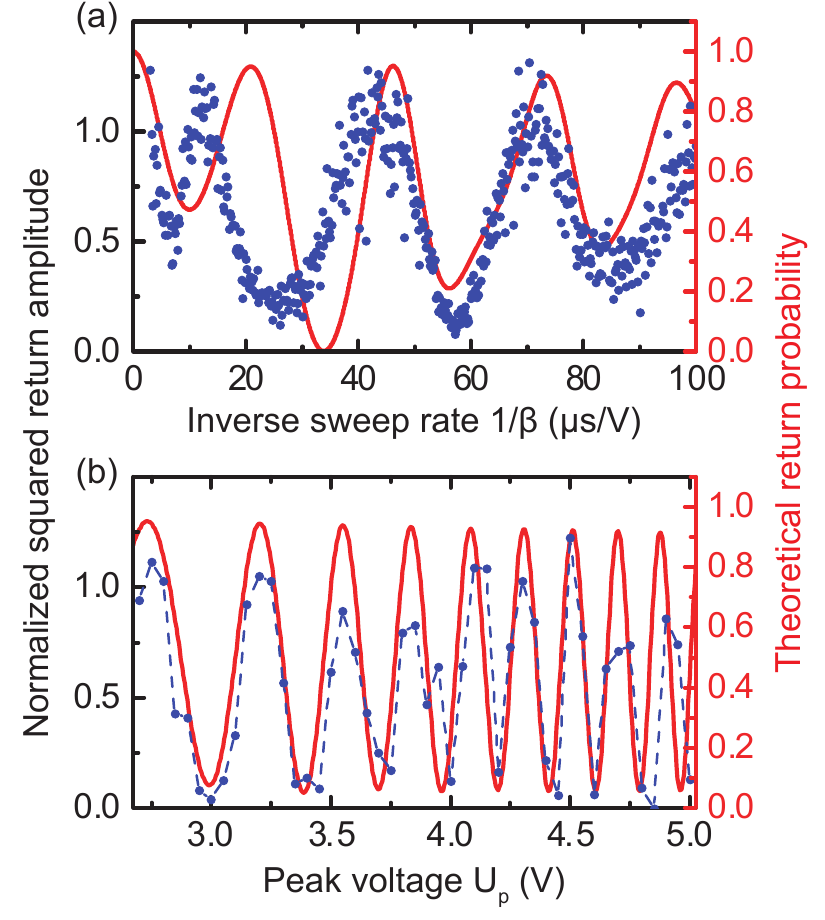}
\caption{\label{fig4}Exemplary classical St\"uckelberg oscillations of sample\,B. Line-cuts of Fig.\,\ref{fig3}\,(a) and Fig.\,\ref{fig3}\,(b). (a) Normalized squared return amplitude (left axis, blue dots) and theoretically calculated return probability given by Eq.~\eqref{eq:cj} (right axis, red line) versus inverse sweep rate for a fixed peak voltage $U_\mathrm{p}=3.3$\,V. (b) Same quantities as above but plotted as a function of peak voltage for a fixed inverse sweep rate $1/\beta=51.6$\,\textmu s/V. Blue dots are joined by blue dashed lines for illustration reasons.}
\end{figure}
\section{Conclusion and Outlook}
\label{conclusion}
In conclusion, we have demonstrated classical St\"uckelberg
oscillations which have previously been experimentally observed exclusively in the framework of
quantum mechanics\,\cite{2010_Shevchenko_Review,2010_Petta_Science,Gaudreau2012,2011_Sun_PRB}. Providing an exact solution for the St\"uckelberg problem\,\cite{1932_Stuckelberg}, we
have established the analogy between the quantum mechanical and the classical return probability.
In this way, we have demonstrated that the coherent exchange of energy between two strongly
coupled classical nanomechanical resonator modes follows the same dynamics as the exchange
of excitations in a quantum mechanical two-level system in the framework of
St\"uckelberg interferometry. However, this analogy generally breaks down if the two coupled harmonic
oscillators start to enter the quantum regime due to multilevel population transfer
effects\,\cite{1997_Usuki_PRB}. This aspect will allow for the future investigation of
dissimilarities, homologies and analogies of classical and quantum mechanical systems as recently
studied in ultracold atoms\,\cite{2016_Lohse_NatPhys,2016_Kaufman_arxiv,2016_Neuzner_NatPhot}. Overall, we have found
remarkably good agreement between experiment and theory. However, parameter regimes yielding larger
deviations are reminiscent of the sensitivity of the exact St\"uckelberg solution to the initial
system parameters, such as the position of the avoided crossing, and hence to fluctuations in the
system. This circumstance, in turn, might be exploited for future investigations in resonator
metrology of decoherence and noise, adapting the approach to employ St\"uckelberg interferometry to
characterize the coherence of a qubit\,\cite{2014_Forster_PRL}. Furthermore, the possibility to
create a superposition state of two mechanical modes may allow for future application as highly
sensitive nanomechanical interferometers\,\cite{2016_Xu_arxiv,2016_Rossi_arxiv,2016_Mercier_arxiv} analogous to
the applications with cold atom and molecule matter-wave
interferometers\,\cite{1997_Andrews_Science,1997_Castin_PRA,2007_Mark_PRL,2011_Kohstall_NJP},
whereas the presence and implications of, e.g., phase noise\,\cite{2015_Maillet_arxiv} can be
resolved by a change in resonator population and interference pattern. Classical St\"uckelberg
interferometry should not be limited to the presented strongly coupled, high quality factor
nanomechanical string resonator modes\,\cite{2013_Faust_NatPhys}, but can in principle be observed in
every classical two-mode system exhibiting the possibility of a double passage through an avoided
crossing within the classical coherence time.\\
%
%

%
%
\begin{appendix}
\renewcommand\thefigure{\thesection.\arabic{figure}}
\section{The nanoelectromechanical system}
\label{sec:nems}
\setcounter{figure}{0}
The nanomechanical device and experimental set-up are depicted in Fig.\,\ref{setup}. The sample investigated at a temperature of 10\,K (sample\,A) consists of a 50\,\textmu m long, 270\,nm wide and 100\,nm thick doubly clamped silicon nitride (SiN) string resonator. The room temperature measurements were conducted on a similar sample (sample\,B), differing only in its resonator length of 55\,\textmu m. As stated in the main text, the temperature does not affect the purely classical character of the system. The string resonators exhibit a high intrinsic tensile pre-stress of $\sigma_\mathrm{SiN}=1.46$\,GPa resulting from the LPCVD deposition of the SiN film on the fused silica substrate. This high stress translates into large intrinsic mechanical quality factors of up to $Q\approx 500,000$, which reduce quadratically with the applied dc tuning voltage in the experiment as a result of dielectric damping\,\cite{2012_Rieger_APL}. Dielectric drive, detection and control are provided via two adjacent gold electrodes in an all integrated microwave cavity enhanced transduction scheme\,\cite{2012_Faust_NatComm,2012_Faust_PRL,2012_Rieger_APL,2013_Faust_NatPhys}.
\begin{figure}[!t]
\includegraphics{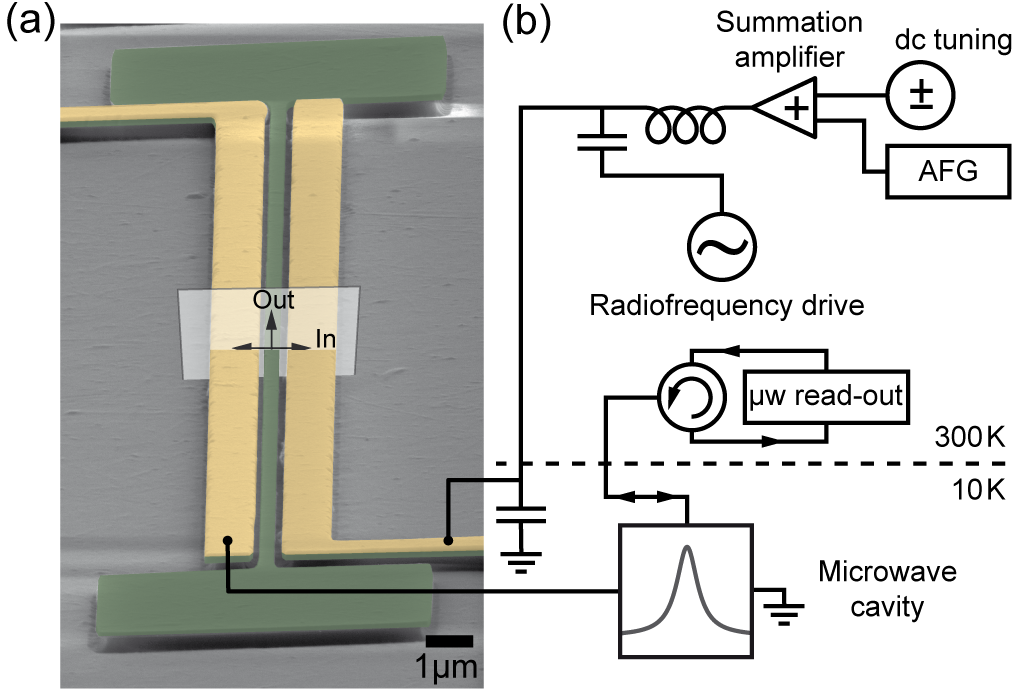}
\caption{\label{setup}Nanoelectromechanical system. (a) False color scanning electron micrograph of the 50\,\textmu m long, 270\,nm wide and 100\,nm thick silicon nitride string (green) in oblique view. The adjacent 1\,\textmu m wide gold electrodes (yellow) are processed on top of the silicon nitride layer. Arrows indicate the flexural mode polarizations out-of-plane (Out) and in-plane (In). (b) Electrical transduction set-up. The arbitrary function generator (AFG) ramp voltage and the dc tuning voltage are added via a summation amplifier and then combined with the rf drive using a bias tee. The microwave read-out is bypassed by the second capacitor, acting as ground path for the microwave cavity.}
\end{figure}
In the experiment, we consider the two orthogonally polarized fundamental flexural modes of the nanomechanical string resonator, namely the oscillation perpendicular to the sample plane (out-of-plane) and the oscillation parallel to the sample plane (in-plane). Applying a dc voltage to one of the two gold electrodes induces an electric polarization in the silicon nitride string resonator, which couples to the field gradient of the inhomogeneous electric field. Consequently, the mechanical resonance frequencies tune quadratically with the applied dc voltage as depicted in Fig.\,\ref{tuning}. Whereas the out-of-plane resonance (Out) tunes towards higher resonance frequencies as a function of dc voltage, the resonance frequency of the in-plane mode (In) decreases\,\cite{2012_Rieger_APL}. Dielectrical tuning of both modes into resonance reveals a pronounced avoided crossing originating from the strong mutual coupling induced by the inhomogeneous electric field. In the coupling region, the mechanical modes hybridize into diagonally ($\pm$45\textdegree) polarized eigenmodes of the strongly coupled system.
\begin{figure}[!htb]
\includegraphics{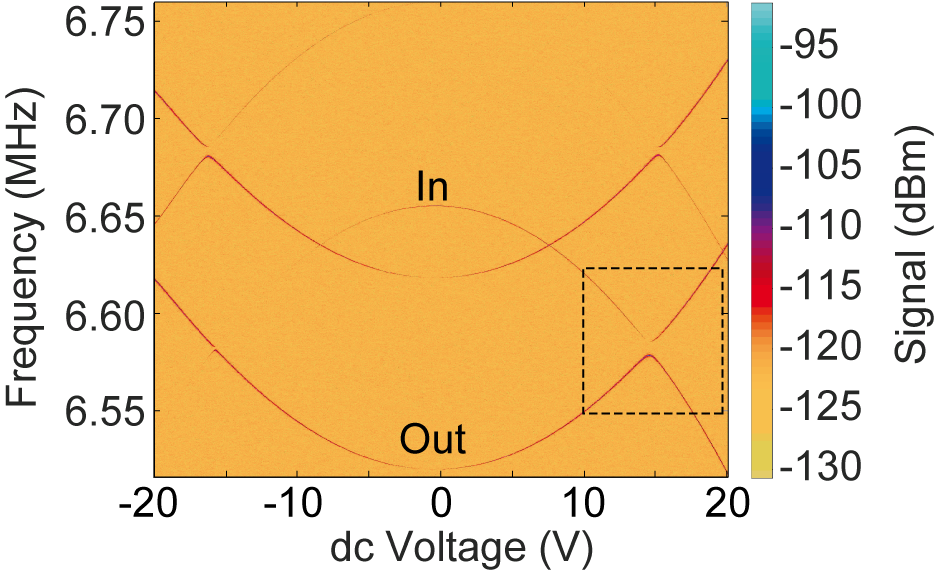}
\caption{\label{tuning}Dielectric frequency tuning. Color-coded frequency spectrum of sample\,B as a function of applied dc tuning voltage. The resonance frequency of the 55\,\textmu m long resonator's fundamental out-of-plane oscillation (Out) increases quadratically as a function of dc voltage. The resonance frequency of the corresponding in-plane mode (In) decreases quadratically. Tuning both modes into resonance, they exhibit a pronounced avoided crossing indicated by the black dashed rectangle. This particular region is displayed in Fig.\,\ref{conv}\,(a). The additional resonances in the spectrum originate from a different mechanical resonator which is coupled to the same microwave cavity.}
\end{figure}
\section{Theoretical Model}
\label{sec:theo}
\setcounter{figure}{0}
In this section, we derive an exact expression for the classical return probability. A detailed
discussion and comparison of our theoretical approach to previous models will be published
elsewhere\,\cite{2016_Seitner_arxiv}.
\\ 
We start by solving the system of first-order differential equations defined in Eq.\,\eqref{eq:classicalflowLZSM} of the main text. Since these equations are formally identical to the Schr\"odinger equation for the (quantum)
Landau-Zener problem, we can follow the work of Vitanov \emph{et al.}~\,\cite{1996_Vitanov_PRA} to
derive the classical flow, $\mathbf{\tilde{c}}(\tau) = \tilde{\varphi} (\tau, \tau_\ui)
\mathbf{\tilde{c}}(\tau_\ui)$ with $\mathbf{\tilde{c}}(\tau) = (\tilde{c}_1 (\tau)\,\, \tilde{c}_2 (\tau))^{\mm{T}}$. Here,
$\tau = \sqrt{\alpha} t$ is a dimensionless time and $\tau_\ui$ the initial dimensionless time. Note
that we use dimensionless times in this chapter in order to provide a derivation which is consistent
with the work of Vitanov \emph{et al.}~\,\cite{1996_Vitanov_PRA}. The equations in dependence of
times in the main text can be recovered by replacement of the dimensionless times following the
above definition. In appendix\,\ref{sec:conv}, we provide the explicit conversion from experimentally accessible
parameters to the dimensionless times. We find 
\begin{equation}
	\begin{pmatrix}
		\tilde{c}_1(\tau) \\
		\tilde{c}_2(\tau) 
	\end{pmatrix}=
	\begin{pmatrix}
		\tilde{\varphi}_{11}(\tau,\tau_\ui) && \tilde{\varphi}_{12}(\tau,\tau_\ui) \\
		-\tilde{\varphi}_{12}^\ast (\tau,\tau_\ui) && \tilde{\varphi}_{11}^\ast (\tau,\tau_\ui) 
	\end{pmatrix}
 	\begin{pmatrix}
		\tilde{c}_1(\tau_\ui) \\
		\tilde{c}_2(\tau_\ui), 
	\end{pmatrix}
	\label{eq:forward_flow}
\end{equation}
with 
\begin{equation}
\begin{aligned}
&\tilde{\varphi}_{11}(\tau, \tau_\ui) =\\
 &\frac{\Gamma\left(1+i\frac{\eta^2}{4} \right)}{\sqrt{2\pi}}
\left[D_{-1-i\frac{ \eta^2}{4}}\left(\mathrm{e}^{-i\frac{3\pi}{4}}\tau_\ui\right)
D_{-i\frac{ \eta^2}{4}}\left(\mathrm{e}^{i\frac{\pi}{4}}\tau\right)\right.\\
&\phantom{=}+\left. D_{-1-i\frac{ \eta^2}{4}}\left(\mathrm{e}^{i\frac{\pi}{4}}\tau_\ui\right)
D_{-i\frac{ \eta^2}{4}}\left(\mathrm{e}^{-i\frac{3\pi}{4}}\tau\right)\right],
\end{aligned}
	\label{eq:phi11}
\end{equation}
and
\begin{equation}
\begin{aligned}
&\tilde{\varphi}_{12}(\tau, \tau_\ui)=\\
&\frac{\Gamma\left(1+i\frac{\eta^2}{4} \right)}{\sqrt{2\pi}}\frac{2}{\eta}
\mathrm{e}^{-i\frac{\pi}{4}}
\left[D_{-i\frac{ \eta^2}{4}}\left(\mathrm{e}^{-i\frac{3\pi}{4}}\tau_\ui\right)
D_{-i\frac{ \eta^2}{4}}\left(\mathrm{e}^{i\frac{\pi}{4}}\tau\right)\right.\\
&\phantom{=}-\left.D_{-i\frac{ \eta^2}{4}}\left(\mathrm{e}^{i\frac{\pi}{4}}\tau_\ui\right)
D_{-i\frac{ \eta^2}{4}}\left(\mathrm{e}^{-i\frac{3\pi}{4}}\tau\right)\right].
	\label{eq:phi12}
\end{aligned}
\end{equation}
Here, $\eta = \lambda/\sqrt{\alpha}$ is the dimensionless coupling, $\Gamma(z)$ is the Gamma
function, and $D_\nu (z)$ is the parabolic cylinder function. To find the flow describing the
evolution of the amplitudes defined in Eq.~\eqref{eq:amp_c1c2}, we apply the unitary transformation defined in
the main text, $\varphi (\tau, \tau_\ui) = S(\tau) \tilde{\varphi} (\tau, \tau_\ui) S^\dag (\tau_\ui)$,
with 
\begin{equation}
	S(\tau) = \exp\left[\frac{i}{4} \tau^2\right] \mathbbm{1}_2.
	\label{eq:S}
\end{equation}
We find 
\begin{equation}
	\varphi (\tau, \tau_\ui) = \exp\left[\frac{i}{4} (\tau^2 - \tau_\ui^2)\right] \tilde{\varphi} (\tau,
	\tau_\ui).
	\label{eq:forward_flow_orig}
\end{equation}
The flow $\varphi (\tau, \tau_\ui)$ describes the evolution of the normalized amplitudes for a
forward sweep; the frequency of mode $1$ ($2$) increases (decreases) with time. This implies that
the back sweep cannot be described by $\varphi(\tau, \tau_\ui)$ since during the evolution the
frequency of mode $1$ ($2$) decreases (increases). Hence, the system of coupled differential
equations describing the dynamics during the backward sweep (denoted by index "b") is given by
\begin{equation}
	i\begin{pmatrix} 
		\dot{\tilde{c}}_\mathrm{1,b} \\ 
		\dot{\tilde{c}}_\mathrm{2,b}
	\end{pmatrix}=
	\begin{pmatrix}
		-\frac{\alpha t}{2} && \frac{\lambda}{2}\\
		\frac{\lambda}{2} && \frac{\alpha t}{2}
	\end{pmatrix}
	\begin{pmatrix}
		\tilde{c}_\mathrm{1,b} \\
		\tilde{c}_\mathrm{2,b} 
	\end{pmatrix}.
	\label{eq:c1c2_matrix_lzsm_back}
\end{equation}
The solutions of Eq.\,\eqref{eq:c1c2_matrix_lzsm_back} can be obtained analogously to
the forward flow since the matrices appearing in Eq.\,\eqref{eq:c1c2_matrix_lzsm} of the main text and
Eq.\,\eqref{eq:c1c2_matrix_lzsm_back} are related by a unitary transformation. We find
\begin{equation}
	\begin{aligned}
		\tilde{\varphi}_\ub (\tau, \tau_\ui) &= \sigma_x \tilde{\varphi} (\tau, \tau_\ui) \sigma_x \\
	&=
	\begin{pmatrix}
		\tilde{\varphi}_{11}^\ast (\tau, \tau_\ui) && - \tilde{\varphi}_{12}^\ast (\tau, \tau_\ui)\\
		\tilde{\varphi}_{12}(\tau, \tau_\ui) && \tilde{\varphi}_{11} (\tau, \tau_\ui)
	\end{pmatrix},
	\end{aligned}
	\label{eq:back_flow}
\end{equation}
where $\sigma_x$ denotes the Pauli matrix in the $x$-direction. The flow describing the evolution
of the amplitudes $c_1 (t)$ and $c_2 (t)$ during the back sweep is obtained as previously, we have 
\begin{equation}
	\varphi_\ub (\tau, \tau_\ui) = \exp\left[\frac{i}{4} (\tau^2 - \tau_\ui^2)\right]
	\tilde{\varphi}_\ub (\tau, \tau_\ui).
	\label{eq:back_flow_orig}
\end{equation}
The state of the system after a double sweep is given by 
\begin{equation}
	\mathbf{c} (\tau) = \varphi_\ub (\tau, -\tau_\up) 
	\varphi (\tau_\up, \tau_\ui) \mathbf{c} (\tau_\ui),
	\label{eq:double_sweep}
\end{equation}
where $\tau_\up$ labels the time at which the first sweep stops and  $-\tau_\up$ corresponds to the
initial time of the back sweep (cf. Eq.~\eqref{eq:solstueckelberg} of the main text). 
As stated in Eq.~\eqref{eq:returnprob} of the main text, the probability to return to mode $1$ is then given by 
\begin{equation}
	\begin{aligned}
	&P_{1\to1} (\tau, \tau_\up, \tau_\ui)=\\ 
	&\abs{\varphi_{11} (\tau_\up, \tau_\ui)
	\varphi_{11}^\ast (\tau, -\tau_\up) + \varphi_{12}^\ast (\tau_\up, \tau_\ui)
	\varphi_{12}^\ast(\tau, -\tau_\up)}^2.
	\end{aligned}
	\label{eq:return_prob_11}
\end{equation}
\section{Conversion factor calibration}
\label{sec:conv}
\setcounter{figure}{0}
\begin{figure*}[!htb]
\includegraphics{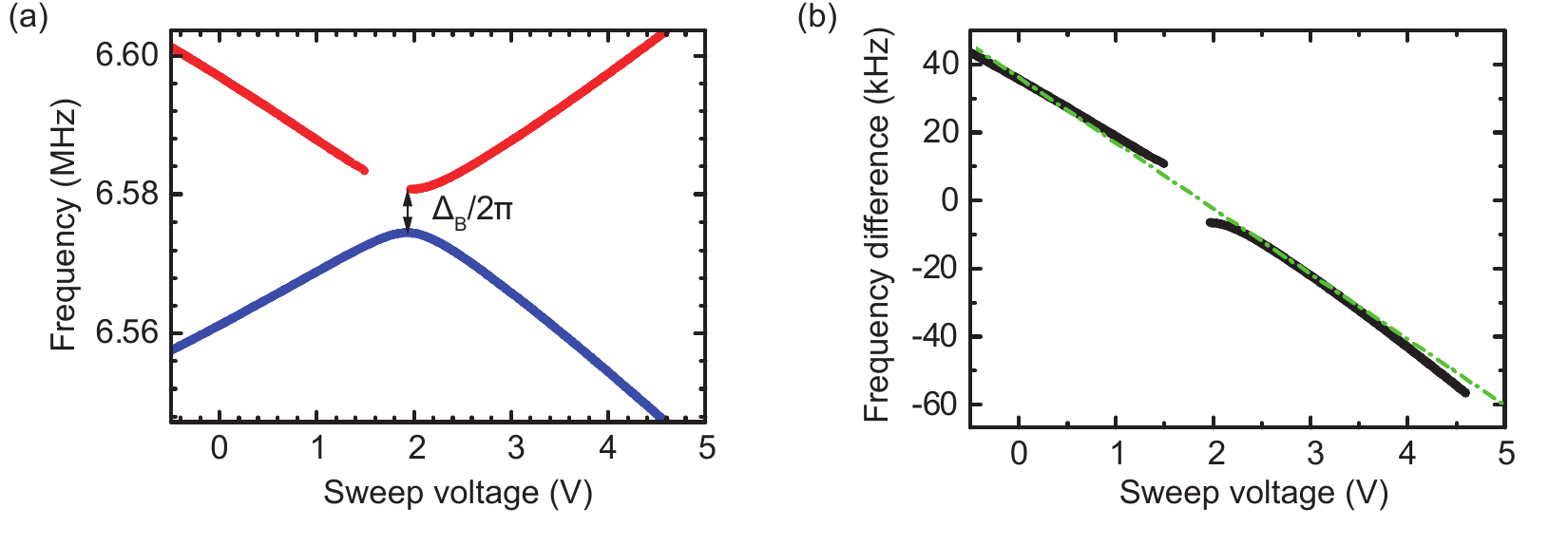}
\caption{\label{conv}Calibration of the conversion factor. (a) Avoided crossing region of sample\,B. A sweep voltage equal to zero corresponds to the initialization point $U_\mathrm{i}=10.4$\,V. The two modes exhibit a frequency splitting of $\Delta_\mathrm{B}/2\pi=6.322$\,kHz at the avoided crossing voltage $U_\mathrm{a}=U_\mathrm{i}+1.958$\,V$=12.358$\,V. The gap in the upper branch (red) results from a signal detection efficiency of the particular mode polarization below the noise level. (b) Frequency difference of the two modes (black) and averaged slope of the linearized frequency tuning illustrated by green dash-dotted line.}
\end{figure*}
In the theoretical model, the state of the system after a double passage through the avoided crossing depends on characteristic sweep times. Experimentally, we realize this double passage by the application of fast triangular voltage ramps, tuning the resonant frequency of the mechanical modes\,\cite{2012_Rieger_APL}. In the following, we focus on sample\,B to illustrate how the different times are obtained.\\
We initialize the resonance in the lower frequency branch at the voltage $U_\mathrm{i}=10.4$\,V, where we apply a continuous sinusoidal drive tone at $\omega_\mathrm{1}(U_\mathrm{i})/2\pi=6.561$\,MHz. We then ramp the sweep voltage up to the peak voltage $U_\mathrm{p}$ across the avoided crossing at voltage $U_\mathrm{a}=U_\mathrm{i}+1.958$\,V$=12.358$\,V and then back to the read-out voltage $U_\mathrm{f}=U_\mathrm{i}+0.5$\,V$=10.9$\,V, where the oscillation energy is read-out again in the lower frequency branch. The offset of the read-out voltage with respect to the initialization voltage is necessary since we cannot stop the sinusoidal drive tone at $\omega_\mathrm{1}(U_\mathrm{i})/2\pi$ during the experiment. For a fixed peak voltage $U_\mathrm{p}$ the voltage sweep is performed for different voltage sweep rates $\beta$, given in the experimental units $[\beta]=\,$V/s. In the theoretical model, the frequency difference of the two modes in units of 2$\pi$ is approximated by $\omega_2-\omega_1\simeq \alpha t$, where the sweep rate $\alpha$ has the dimensions $[\alpha]=2\pi\times$\,Hz/s. Consequently, we introduce the conversion factor $\zeta$ from voltage to frequency, defined via the relation
\begin{equation}
\label{conveq}
\alpha=2\pi\times\zeta\beta.
\end{equation}
Figure\,\ref{conv} illustrates the calibration of the conversion factor. As conventional in experiments on St\"uckelberg interferometry, the frequency difference of the two mechanical modes is approximated to be linear in time, i.e. linear in sweep voltage. In our particular system the resonance frequencies of the mechanical flexural modes tune quadratically with voltage outside of the avoided crossing (see Fig.\,\ref{tuning}). Nevertheless, for the designated region around the avoided crossing, the two frequency branches can be linearized as follows. We take the frequency difference of both modes before and after the avoided crossing (cf. Fig.\,\ref{conv}\,(b)), respectively, and extract the slopes via a linear fit. The two different slopes on the left and the right hand side of the avoided crossing are averaged, yielding an effective conversion factor (dash-dotted green line) 
\begin{equation}
\zeta=19.224\,\mathrm{\frac{kHz}{V}}.
\end{equation}
Depending on the specific peak voltage $U_\mathrm{p}$, one could take into account a weighted average of the two slopes in order to mitigate the deviation of the quadratic frequency tuning from the linear approximation. Here, one has to point out deliberately that we neglect any weighted average, but take solely the above conversion factor for the calculation of the theoretical return probabilities. We are well aware of the fact that this linearization translates into a direct discrepancy between the theoretical model and the experimental results. Nevertheless, in our opinion, these discrepancies are prevailed by the benefits of a closed theoretical calculation using a single set of parameters which is supported by the remarkably good agreement between experiment and theory. Hence, we express the characteristic sweep times in the theoretical model by the following parameters extracted from the avoided crossing in Fig.\,\ref{conv}\,(a):
\begin{eqnarray}
\label{conv_time}
\begin{aligned}
&t_\mathrm{i}=-\frac{1}{\beta}(U_\mathrm{a}- U_\mathrm{i})=\frac{\tau_\mathrm{i}}{\sqrt{\alpha}}\\
&t_\mathrm{p}=\frac{1}{\beta}(\widetilde{U}_\mathrm{p}- U_\mathrm{a})=\frac{\tau_\mathrm{p}}{\sqrt{\alpha}}\\
&t_\mathrm{f}=\frac{1}{\beta}(U_\mathrm{a}- U_\mathrm{f})=\frac{\tau_\mathrm{f}}{\sqrt{\alpha}},
\end{aligned}
\end{eqnarray}
where $\widetilde{U}_\mathrm{p}=U_\mathrm{i}+U_\mathrm{p}$. As explained above, the return probability is measured at the read-out voltage $U_\mathrm{f}\neq U_\mathrm{i}$. Consequently, we replace $\tau$ by $\tau_\mathrm{f}$ in the back sweep of the theory, which modifies Eq.\,\eqref{eq:return_prob_11} to
\begin{equation}
	\begin{aligned}
	&P_{1\to1} (\tau_\mathrm{f}, \tau_\up, \tau_\ui)=\\ 
	&\abs{\varphi_{11} (\tau_\up, \tau_\ui)
	\varphi_{11}^\ast (\tau_\mathrm{f}, -\tau_\up) + \varphi_{12}^\ast (\tau_\up, \tau_\ui)
	\varphi_{12}^\ast(\tau_\mathrm{f}, -\tau_\up)}^2.
	\end{aligned}
	\label{eq:return_prob_11_mod}
\end{equation}

\section{Temperature fluctuations}
\label{sec:temp}
\setcounter{figure}{0}
As stated in the main text, the measurement of the normalized squared return amplitude for various voltage sweep rates $\beta$ at a particular peak voltage $U_\mathrm{p}$ takes up to 16\,hours. During this time, the ambient temperature undergoes fluctuations of $\pm 2$\,K per hour due to insufficient air conditioning. Since the mechanical resonance frequency shifts due to thermal expansion of the silicon nitride by approximately 500\,Hz/K, both resonances shift by approximately 40\,linewidths. In order to initialize the system at the same resonance frequency for every particular measurement, we implement a feedback loop which regulates the initialization voltage. Therefore, the initialization voltage slightly shifts from measurement to measurement, reflecting the temperature fluctuations. Figure\,\ref{fluct} depicts the initialization voltage shift versus inverse sweep rate for the dataset of peak voltage $U_\mathrm{p}=3.3$\,V, which corresponds to the measurement depicted in Fig.\,\ref{fig4}\,(a) of the main text. Each point represents a single measurement for a particular sweep rate. The first measurement is performed at an inverse sweep rate of 100\,\textmu s/V at the initialization voltage $U_\mathrm{i}=10.4$\,V and therefore corresponds to a shift of zero volts. Clearly, the temperature fluctuations not only affect the initialization voltage required to obtain the desired resonance frequency, but will also alter other system parameters, such as the position of the avoided crossing $U_\mathrm{a}$, that greatly affect the theory (cf. appendix\,\ref{sec:conv}). Consequently, the temperature fluctuations lead to deviations between experiment and theory, since we calculate the return probability with a single set of parameters. In turn, these deviations might be used to infer fluctuations of the system in future applications of St\"uckelberg interferometry.
\begin{figure}[!htb]
\includegraphics{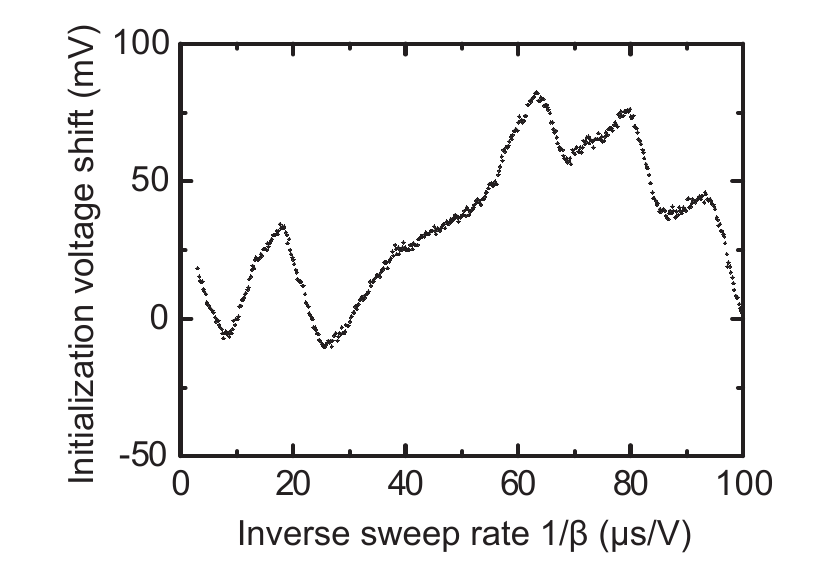}
\caption{\label{fluct}Temperature fluctuations. Initialization voltage shift versus inverse sweep rate for the dataset depicted in Fig.\,\ref{fig4}\,(a) ($U_\mathrm{p}=3.3$\,V) of the main text. Each point corresponds to the measurement of the normalized squared return amplitude for a given inverse sweep rate. The first measurement is performed at 1/$\beta=100$\,\textmu s/V, representing the initialization at resonance frequency $\omega_\mathrm{i}(U_\mathrm{i})/2\pi$ for voltage $U_\ui=10.4$\,V. The implemented feedback loop
regulates the initialization voltage in order to compensate the temperature fluctuations of the
mechanical resonance. Consequently, the voltage shift illustrates the fluctuations of the ambient
temperature.}
\end{figure}
\section{Experimental uncertainties}
\label{sec:fluc}
\setcounter{figure}{0}
In Fig.\,\ref{linecut} we provide additional horizontal and vertical line-cuts from Fig.\,\ref{fig3} of the main text. We observe pronounced oscillations in the normalized squared return amplitude (blue dots) as well as in the theoretically calculated return probability (red line). Nevertheless, the deviations between experiment and theory are more apparent, especially for Fig.\,\ref{linecut}\,(b), which depicts a vertical line-cut for a fixed inverse sweep rate of $1/\beta=60$\,\textmu s/V, i.e. within the "plateau" in Fig.\,\ref{fig3}\,(b) of the main text. Whereas the normalized squared return amplitude exhibits destructive interference, with the signal dropping close to zero, the minima in the return probability saturate at a value of approximately 0.3. This discrepancy is supposed to originate from the high sensitivity of the theoretical model to the input parameters. Experimental uncertainties and fluctuations deter the system from interference with the same constant parameters throughout all individual measurements. Since the "plateau" in the theory is characteristic for a particular set of exact and constant parameters, it cannot be recovered under the given experimental conditions.\\
\begin{figure}[!htb]
\includegraphics{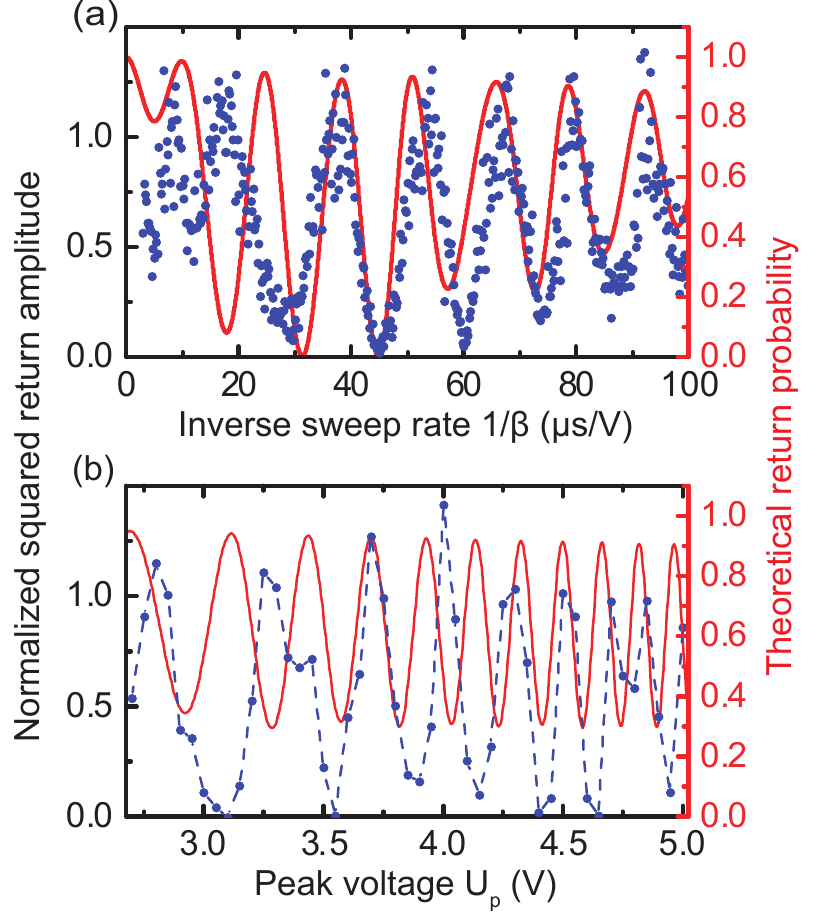}
\caption{\label{linecut}Classical St\"uckelberg oscillations. (a) Normalized squared return amplitude (left axis, blue dots) and theoretically calculated return probability given by Eq.~\eqref{eq:cj}\, of the main text (right axis, red line) versus inverse sweep rate for a fixed peak voltage $U_\mathrm{p}=3.85$\,V. (b) Same quantities as above but plotted as a function of peak voltage for a fixed inverse sweep rate $1/\beta=60$\,\textmu s/V. Blue dots are joined by blue dashed lines for illustration reasons.}
\end{figure}
The experimental uncertainties arise not solely from the temperature fluctuations. The voltage ramp also affects the characteristic parameters, such as the exact position of the avoided crossing $U_{\mathrm{a}}$. As previously stated, the dc voltage induces dipoles in the silicon nitride string resonator, which couple to the electric field gradient. A variation in dc voltage changes the inhomogeneous electric field at the same time, to which the nanoelectromechanical system needs to equilibrate. Consequently, the resonance frequencies of the mechanical modes drift towards the equilibrium position of the system. This drift, in turn, alters the characteristic system parameters, i.e. the characteristic voltages used for the theoretical calculations, and depends on the magnitude of the peak voltage $U_\mathrm{p}$. Concerning the initialization voltage, we simultaneously account for this effect via the initialization feedback loop (see section IV). Nevertheless, the exact position of the avoided crossing $U_\mathrm{a}$ varies slightly due to this retardation effect. Experimentally, we mitigate the influence of this drift by means of a "thermalization" break of 10 seconds after each voltage ramp.\\
Another possible uncertainty arises from the imprecision in the value of the peak voltage $U_\mathrm{p}$ at the sample. The output amplitude uncertainty of the arbitrary function generator used in the room temperature experiments is classified by the manufacturer as $\pm$1\,\% of the nominal output voltage. Consequently, the maximum uncertainty in the peak voltage corresponds to $\pm$0.05\,V for a maximum peak voltage of $U_\mathrm{p}=5.0$\,V, which is equal to the voltage step size between two horizontal lines of Fig.\,\ref{fig3}\,(a) in the main text.\\
As stated in the main text, we observed additional deviations in the experimental data of sample\,B from the theory for very fast voltage sweeps ($1/\beta \leq 50$\,\textmu s/V). These deviations originate from a flattening of the triangular voltage ramps. Records of the triangular voltage pulse taken by an oscilloscope revealed a flattening of the voltage apex depending on the peak voltage $U_\mathrm{p}$, which becomes significant for very fast sweeps. This flattening translates into a peak voltage cut-off and hence a different value of $U_\mathrm{p}$, which is transduced to the sample. We attribute this to the limited bandwidth of the summation amplifier, which reduces the pulse fidelity for very short ramp times. In the experiments conducted on sample\,A, a high performance summation amplifier has been employed together with a different arbitrary function generator. The latter exhibits a greatly enhanced bandwidth and sampling rate (nearly one order of magnitude) compared to the device employed in the room temperature experiment. As a consequence, the flattening of the voltage pulse apex is less pronounced and we find good agreement between the experimental data and the theory for inverse voltage sweep rates $1/\beta \leq 50$\,\textmu s/V.\\
\end{appendix}
%
%
%

\begin{acknowledgements}
Financial support by the Deutsche Forschungsgemeinschaft via the collaborative research center SFB 767 and via project Ko\,416-18 is gratefully acknowledged. H.\,R. acknowledges funding from the Swiss SNF. We thank Aashish A. Clerk for critically reading the manuscript.
\end{acknowledgements}

\end{document}